\documentclass[apj]{emulateapj}
\usepackage{natbib}
\usepackage{graphicx}

\shorttitle{Diffusion in structured turbulence}
\shortauthors{Laitinen et al.}

\begin{document}

\title{ENERGETIC PARTICLE DIFFUSION IN STRUCTURED TURBULENCE}

\author{T. Laitinen, S. Dalla, and J. Kelly}
\affil{Jeremiah Horrocks Institute, University of Central Lancashire,
  PR1 2HE Preston, UK}

\begin{abstract}
  In the full-orbit particle simulations of energetic particle
  transport in plasmas, the plasma turbulence is typically described
  as a homogeneous superposition of linear Fourier modes. The
  turbulence evolution is, however, typically a nonlinear process,
  and, particularly in the heliospheric context, the solar wind plasma
  is inhomogeneous due to the transient structures, as
  observed by remote and in-situ measurements. In this work, we study
  the effects of the inhomogeneities on energetic particle transport
  by using spatially distributed, superposed turbulence
  envelopes. We find that the cross-field transport is significantly
  reduced, when compared to the results obtained with homogeneous
  turbulence. The reduction can reach an order of magnitude when the
  enveloping breaks the wave phase coherence along the mean magnetic
  field direction.
\end{abstract}


\section{Introduction}\label{sec:introduction}

When studying solar energetic particle (SEP) events, it is crucial to
uncover the propagation of the particles in the interplanetary
medium. Energetic particles related to solar eruptions are observed at
large range of latitudes and longitudes
\citep[e.g.,][]{DallaEa2003Annales,Liu2011}, and without understanding
the propagation of particles across the mean magnetic field it is
difficult to determine the extent of the acceleration region, and its
relation to the physics of the eruption. 

The modelling of cosmic ray transport in turbulent plasmas is
typically based on the diffusion-convection equation
\citep{Parker1965}, and the quasilinear approach incorporating pitch
angle diffusion description, with the cross-field propagation
described as field line random walk \citep{Jokipii1966}. The modelling
of pitch angle evolution has evolved to include adiabatic focusing
\citep{Earl1976}, adiabatic deceleration
\citep{Ruffolo1995,Kocharov1998} and structured solar wind
\citep{Kocharov2009}. The application of these models to obtaining
the SEP injection close to the Sun from the observations at 1 AU
mostly concentrated on the particle propagation along the mean
magnetic field \citep[e.g.][]{Laitinen2000,Droge2003}. Only recently
has the cross-field propagation been included into the SEP studies
\citep[e.g.][]{Zhang2009,Droge2010,He2011}.

A wide range of values for the perpendicular diffusion coefficient
have been reported. Studies of the transport of galactic cosmic rays
\citep[e.g.][]{Burger2000} and Jovian electrons \citep[][and
references therein]{Zhang2007} give estimates of order
$\kappa_\perp/\kappa_\parallel=0.01$. On the other hand,
\citet{Zhang2003} report values of
$\kappa_\perp/\kappa_\parallel=0.25$ in relation to protons
originating from a solar event, while \citet{DwyerEa1997} find the
ratio to be of order of unity for energetic particles
observed in association with corotating interaction regions.

The values of $\kappa_\perp/\kappa_\parallel$ obtained from
observations are related to different regions in the heliosphere, and
possibly to different properties of the turbulence, which poses a
problem when comparing with theoretical work. In recent years, the use
of full-orbit simulations of energetic particles in studying particle
propagation has gained attention. The benefit of this approach is that
the parameters defining the turbulence properties can be explicitly
prescribed. In addition, for the energetic particle propagation no
a~priori assumptions, such as the applicability of the diffusion
description, are required. The approach has been used by
\citet{GiaJok1999}, who obtained
$\kappa_\perp/\kappa_\parallel=0.01-0.03$, with the perpendicular
diffusion coefficient contradicting the quasilinear theory results of
\citet{Jokipii1966}. \citet{Qin2002_apjl} used the method to discover
subdiffusion phase in the particle propagation perpendicular to the
mean magnetic field, followed by subsequent recovery of
diffusion. This phenomenon has subsequently been modelled as interplay
between parallel and perpendicular propagation effects
\citep{Matthaeus2003}. Thus, the full-orbit particle simulations have
proved a valuable tool when studying particle transport.

The full-orbit particle simulation results may depend, however, on the
way the turbulent electric and magnetic fields are described. A
popular choice is to superpose Fourier modes on a constant magnetic
field \citep[e.g.][]{GiaJok1999, Qin2002, Qin2002_apjl, Zimbardo2006,
  RuffoloEa2008} or more recently, on a Parker spiral field
\citep{Tautz2011}. In this approach, however, the modes remain linear
throughout the simulation domain, while the heliospheric plasma is
known to evolve non-linearly in addition to linear mode propagation
\citep[see, e.g.,][for a review]{TuMarsch1995}. Thus, the Fourier-mode
description is not consistent with the behaviour of plasma turbulence,
and using the approach may result in inaccurate particle diffusion
coefficients. Some work exists on simulating particles in turbulent
fields obtained by using MHD simulations, where the nonlinear
interactions are properly addressed
\citep[e.g.,][]{Beresnyak2011}. However, such studies are limited by
computational requirements to a small range of scales and to cartesian
geometries with constant background fields.

The assumption that the plasma turbulence is spatially homogeneous, while
useful when comparing the simulations to theoretical models, can also
pose a problem when trying to understand SEP observations. The
heliospheric magnetic field expands as a Parker spiral, and may have a
complicated structure
\citep[e.g.,][]{Parker1958,Fisk1996,Sternal2011}. The solar wind is
permeated by transient structures, as shown by in-situ and
remote-sensing observations \citep{Rouillard2010a,
  Rouillard2010b}, thus making it a spatio-temporally varying medium,
instead of a steady solar wind flow. 
Thus, for SEP analysis there is
a need to understand how such a structured plasma environment
influences energetic particle propagation.

In this work, we study the effect of structures in the turbulence on
energetic particle propagation by means of full-orbit energetic
particle simulations. We describe the turbulence as a superposition
of localised, randomly distributed envelopes that contain a sum of
linear waves. For simplicity, the envelopes are not limited in
  the direction perpendicular to the mean field, but the variation
  takes place only along the mean field direction. We introduce two
different models to study the effects of the structures. In the {\em
  modulated wavefield} model, only the amplitude of the turbulence is
modulated, with the individual Fourier modes remaining coherent
throughout the simulation domain. In the {\em random envelope} model,
the phases, wave normal directions and polarisations of the modes are
different in different envelopes. This is done to mimic the non-linear
evolution of turbulence being convected throughout the
heliosphere. Of these two models, the random envelope model
  captures better the influence of nonlinear interactions present in
  the heliospheric turbulence. Comparing the results of the random
  envelope model with those of the modulated wavefield model
  allows us to distinguish between the effects of amplitude modulation
  on one hand, and the phase, wave normal and polarisation changes on
  the other, as both effects are present in the random envelope
  model.

The turbulence is superposed on a constant background magnetic
  field. While such a background field limits the applicability of the
  model results to the heliospheric environment, it is useful as a
  first approach in isolating the energetic particle transport effects
  due to the structured turbulence. The effects of a spatially
  non-uniform heliospheric magnetic field, and of fully
  three-dimensional structures, will be the subject of future
  studies.

In Section~2 we introduce our description of the structured
turbulence, and our approach to obtain the energetic particle
diffusion coefficients. In Section~3, we present the turbulence
characteristics, and the full-orbit particle simulations and the
derived diffusion coefficients, and study their variation with
envelope characteristics, which can be considered as a measure for the
structure size in the turbulent plasmas. We discuss the implications
of the results and draw our conclusions in Section~4.

\section{Model}\label{sec:model}

\subsection{Turbulence Model}

The turbulence model consists of envelopes containing fluctuating
magnetic field given by a sum of infinite plane waves. The magnetic
field is defined as 
$$
\mathbf{B}=\mathbf{B}_0 \hat e_z + \delta\mathbf{B}(x,y,z), 
$$
where the background magnetic field is uniform and constant, with
$B_0=5$~nT, consistent with the magnetic field at 1~AU. The
fluctuating field $\delta\mathbf{B}$ is based on the homogeneous
turbulence model of \citet{GiaJok1999}, and is defined as a sum of $N$
modes,
\begin{equation}
  \delta \mathbf{B}(x,y,z)=\sum_{n=1}^N A(k_n)
  \hat\mathbf{\xi}_n\exp\{i(k_n z'_n+\beta_n)\}.\label{eq:gjturb}
\end{equation}
Here $\hat\xi_n$ is the polarisation vector,
$$\hat\xi_n=\cos\alpha_n \hat \mathbf{x}'_n+i \sin\alpha_n \hat \mathbf{y}'_n.$$
The unprimed coordinate system has the background magnetic field along
the $z$-axis, and the primed coordinate system is obtained from the
unprimed through rotation characterised by the spherical coordinate
system angles $\theta_n$ and $\phi_n$ of the wave vector
$\mathbf{k}_n$, as defined in \citet{GiaJok1999}. The fluctuation
amplitude $A(k_n)$ is given by a power law spectrum
$$ A^2(k_n) =  B_1^2 \frac{G_n}{\sum_{n=1}^{N} G_n},
 \;\; G(k_n) = \frac{\Delta V_n}{1+(k_n L_c )^\gamma}, 
$$ 
where $B_1^2$ is the variance of the magnetic field, $L_c$ the
spectrum's turnover scale, $\gamma$ the spectral index, and
$\Delta V_n$ specifies the volume element in $k$-space that the discrete
mode $k_n$ represents. For the turnover scale, also called correlation
length, we use $L_c=2.15\, r_\odot$, where $r_\odot$ is the solar
radius.

In our simulations, we use the composite spectrum model, which is
composed of a sum of slab and two-dimensional (2D) turbulence \citep{Gray1996}. For the
slab component, $\mathbf{k}_n$ is directed along the background field,
which is parallel to the $z$-axis, and the polarisation vector
lies in the $xy$-plane, with a random polarisation angle $\alpha_n$. For the
2D-component, $\mathbf{k}_n$ lies in the $xy$-plane. The polarisation
vector is also in the $xy$-plane, with $\alpha_n=\pi/2$, perpendicular
to $\mathbf{k}_n$, to satisfy $\nabla \cdot {\mathbf{B}}=0$.  The
azimuthal angle $\phi_n$ and the random phase $\beta_n$ are chosen
from a uniform random distribution.

The slab and 2D spectra have the same turnover scale, and the
wavemodes are generated for the same wavenumbers $k_n$. The
wavenumbers are logarithmically spaced between $2\pi/1 \mathrm{AU}$
and $2\pi/10^{-4} \mathrm{AU}$. The spectral indices for the slab and
2D components are 5/3 and 8/3, respectively, and the factor $\Delta
V_n$ is $\Delta k_n$ for the slab component and $2\pi k_n\Delta k_n$
for the 2D component. The total turbulence energy is divided between
the slab and 2D component with ratio 20\%:80\%, following
\citet{Bieber1996}.

\subsection{Turbulence Envelope Model}

We envelope the infinite plane waves, given by Eq.~(\ref{eq:gjturb})
into $N_p$ wave packets of length $L_p$, randomly distributed in the
simulation box. The envelopes are of cosine shape, given by
\begin{eqnarray}
 \mathfrak{A}_i(z)&=&\frac{1}{2}\left[1-\cos\left(2\pi
  \frac{z-z_i}{L_{p}}\right)\right]\cdot \label{eq:env_amplitudes} \\
&&\bigg[\Theta\{z-z_i\}-\Theta\{z-(z_i+L_p)\}\bigg],\nonumber
\end{eqnarray}
with $z_i$ the starting point of the envelope, and the Heaviside
functions $\Theta\{z\}$ limit each envelope to the range $z_i<
z<z_i+L_p$. The magnetic field in these envelopes remains
divergence-free, as the envelope amplitude changes only in direction
perpendicular to the polarisation vectors $\hat\xi_n$.  The overall
turbulent magnetic field is the sum of the magnetic field in the
envelopes,
\begin{equation}
 \delta \mathbf{B}(x,y,z)=\sum_{i=1}^{N_p} \mathfrak{A}_i(z)\;\delta\mathbf{B}_{i},
\end{equation}
where $\delta\mathbf{B}_{i}$ is given by
Eq.~(\ref{eq:gjturb}). The distribution and summing of the
  envelopes is shown in the bottom panels of
  Fig.~\ref{fig:amplitude}, where the grey-filled curves show the
  amplitudes of each envelope, and the thick curve their sum.

The envelopes are spread randomly along the $z$-direction from $z=0$
to $z=L_{max}$, where the length of the simulation volume along the
$z$-axis, $L_{max}$, is chosen large enough that the particles will
remain in the volume for the duration of the simulation. As the
variation of the amplitudes takes place only in the $z$-direction,
there is no need to limit the volume in the $xy$-plane. The number of
the envelopes, $N_p$, is determined by defining an envelope density
$\rho_p=N_{p}L_p/L_{max}$, which describes the filling of the
simulation space with the turbulent envelopes.

We consider two different models for $\delta\mathbf{B}_{i}$. In the
{\em wavefield modulation} model, the set of angles $\beta_{n}$,
$\phi_{n}$ and $\alpha_{n}$ are the same for each envelope, and thus
the enveloping only modulates the turbulent field described by
Eq.~(\ref{eq:gjturb}). In the {\em random envelope} model, each
envelope $i$ has a unique set of angles $\beta_{ni}$, $\phi_{ni}$ and
$\alpha_{ni}$, mimicking the spatial evolution of the turbulence
phases and wavenumbers.

\subsection{Turbulence Energy Density}

To compare the turbulent fields and their effects on SEP propagation,
we scale the fluctuation amplitudes so that the average energy density
in the fluctuating field is independent of the enveloping
parameters. As the energy density varies considerably in the enveloped
cases, we calculate the energy density over a large spatial
domain. The average energy density in a volume of cross-section $S$
and height $L_{max}$ is given by
$$
U=\frac{1}{8\pi}\frac{1}{S\; L_{max}}\int\int \delta B^2 dz\;dS=\frac{1}{8\pi}\left<\delta B^2\right>,
$$
where $\left<\delta B^2\right>$ is the variance of the fluctuations over
the volume. The energy density for the enveloped turbulence is then
\begin{equation}
 U_{env} = \frac{1}{8\pi L_{max}}\int_{0}^{L_{max}} 
 \sum_{i=1}^{N_p}\sum_{j=1}^{N_p}\mathfrak{A}_i\mathfrak{A}_j
 \delta \mathbf{B}_i\cdot\delta \mathbf{B}_j  \;dz. \label{eqn:envEdens}
\end{equation}

For the random envelope model, this integral can be estimated
analytically (see Appendix \ref{sec:turbe}). For the wavefield
modulation model, the scaling factor is obtained through numerical
integration.

\subsection{Energetic Particle Simulations}

In order to study charged particle propagation in turbulent magnetic
fields, we integrate the fully relativistic equation of motion of
energetic protons using the simulation code by \citet{Dalla2005}. As
the fluctuating field is magnetostatic, the electric field is
zero. For the integration, we use the Bulirsh-Stoer method
\citep{NumRecp}. The method uses adaptive timestepping to control the
accuracy by limiting the error between consecutive steps to a given
tolerance. We use the tolerance of $10^{-9}$, which ensures the
conservation of energy to within $10^{-6}$ in our simulations. The
particles' initial velocity distribution is isotropic and
monoenergetic, and their initial positions are selected randomly in a
region of size 40 times the turbulence correlation scale, in order to
minimise the effects of local magnetic field. For each set of
turbulence and particle parameters, we run the simulations for 10
field realisations, with 2048 particles for each simulation. In the
turbulence realisations, we typically use $N=128$ Fourier modes, with
validation runs using $N=1024$ modes to verify that the results are
not altered by the number of modes.

The diffusion coefficient is obtained from the definition
\begin{equation}
\kappa_{\zeta\zeta}=\frac{\left<\Delta\zeta^2\right>}{2t},
\,\,\,\,\,\zeta=x,y,z,\label{eq:diffcoeff}
\end{equation}
\citep[e.g.][]{GiaJok1999}, where the square of the displacement,
$\Delta\zeta^2$, is averaged over the simulated particles. The diffusion
coefficients are calculated separately for each field realisation. In
Figs.~\ref{fig:kappapar_ene}--\ref{fig:kappaperp_lencomp}, $\kappa_{par}$ refers to $\kappa_{zz}$, whereas
$\kappa_{perp}$ is obtained as the mean of $\kappa_{xx}$ and
$\kappa_{yy}$. 

The diffusion coefficients experience typically super- and
subdiffusive phases, after which they settle to constant values, which
we use in this study. In practice, we continue the simulations to
$\sim100$ parallel diffusion times in each run. In addition to this,
we require that the particles have spread over a large number of
envelopes. For this, we require that the FWHM of the distribution, as
calculated from the analytic solution to the diffusion equation
\citep[e.g.][]{GiaJok1999}, $4\sqrt{\kappa_{zz} t \ln 2}$ is at least
$40 L_p$. This is done in order to reduce the statistical errors
caused by the local differences between different field realisations.

We have reproduced some of the simulations done by \citet{GiaJok1999},
and found good agreement with their example run presented their
Appendix~A. We found, however, that the agreement between their
experiments~1 and~3 is somewhat dependent on the selection of
simulation domain size and fitting parameters when using the
experiment~1. For this reason, the results presented in this study may
differ somewhat from their results, as they use the experiment~1 for
their results, whereas our method corresponds to their experiment~3.

The diffusion of the particles can be caused by fieldline
  meandering, and by the drifts of the particles relative to the
  fieldlines.  However, these processes are strongly coupled. Even a
  small drift of a particle from its original fieldline can result in
  large deviation from the original fieldline, depending on the local
  structure of the magnetic field. For this reason, separating the
  contributions of the two effects is not trivial. In this study, we
  do not attempt to make this distiction, but consider the diffusion
  as a compound effect of the two processes.

\section{Results}

In this work, we study the effect of turbulent structures on energetic
particle transport by using the turbulence model described in
Section~\ref{sec:model}, varying the enveloping parameters. For the
envelope density $\rho_p$ we have chosen to use values 1,~4 and 16, to
represent tenuous to dense enveloping. The length of the envelope,
$L_p$, is chosen to have values $L_c$, $4\,L_c$ and $16\,L_c$. In this
section, we first study how the enveloping affects the turbulence, and
then how the particle propagation is affected.

\subsection{Enveloped Turbulent Field}

 \begin{figure}
   \plotone{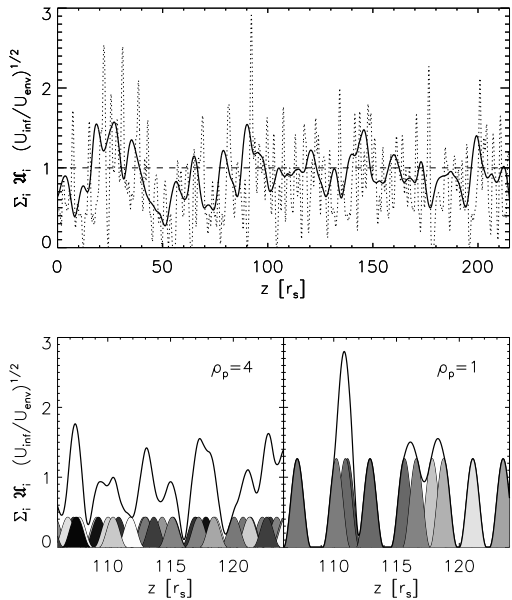}
  \caption{\label{fig:amplitude} {\em Top panel}: The sum of envelope
    amplitudes for parameters $\rho_p=16$ and $L_p=4 L_c$ (solid
    curve), and $\rho_p=4$ and $L_p=1 L_c$ (dotted curve), in the
    modulated wavefield model. The horizontal dashed line represents
    the homogeneous model. Here, the energy density scaling factor
    $4/\rho^2$ has been used. {\em Bottom panels}: The curves filled
    with greyscale colours show the individual envelopes, with
    $L_p=L_c$. The
    thick curve represents the sum of the envelopes.}
 \end{figure}

To understand how the enveloping affects the turbulence, it is useful
to compare the behaviour of the envelope amplitudes to the homogeneous
turbulence.  In the top panel of Fig.~\ref{fig:amplitude} we
plot the sum of the envelope amplitudes for the wavefield modulation
model, for two combinations of $\rho_p$ and $L_p$. As can be seen, the
high-density enveloping, with $\rho_p=16$, results in smaller
deviations from the homogeneous model (the horizontal dashed curve)
than the intermediate density ($\rho_p=4$). Thus, the envelope density
describes the strength of the deviation of the mean amplitude from a
constant value.  The tenuous enveloping, with $\rho_p=1$
(bottom right panel of Fig.~\ref{fig:amplitude}), implying an
  average of one envelope at a given position, results in regions with
  no turbulence at all, and regions with overlapping envelopes. While
  such a situation is not necessarily realistic, it is taken as an
  extreme of the parametrisation, and can in later studies be more
  realistically supplemented with some small underlying homogeneous
  turbulence. At the large density limit, the wavefield modulation
  approaches the homogeneous turbulence description.

 \begin{figure}
   \plotone{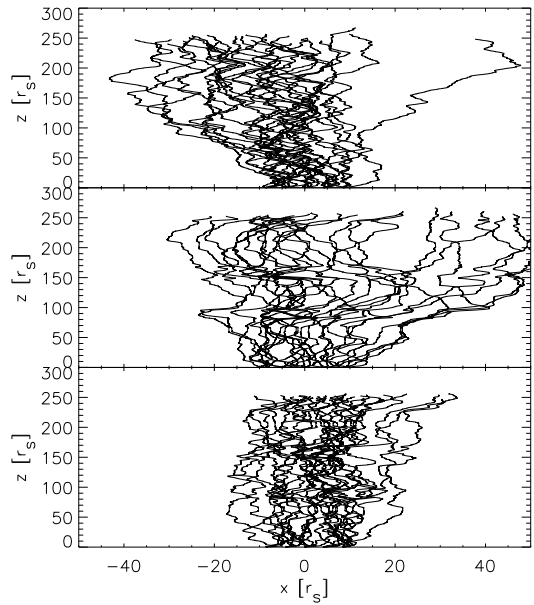}
   \caption{\label{fig:fieldlines} The field lines in the homogeneous
     (top panel), modulated wavefield (middle panel) and the random
     envelope turbulence model (bottom panel). In the latter two
     models, the envelope length is $L_p=4 L_c$ and density
     $\rho_p=4$.}
 \end{figure}

We can also visualise how the field line wandering is affected by the
enveloping. Fig.~\ref{fig:fieldlines}, shows the fieldlines for
homogeneous turbulence (top panel), the modulated wave field (middle
panel) and the random envelope model (bottom panel), for parameters
$L_p=4 L_c$ and $\rho_p=4$. The field lines are initiated from an area
of $10 L_c \times 10 L_c$ at $z=0$. The most striking difference can
be seen in the reduced spreading of the field lines in the case of
random enveloping (bottom panel of Fig.~\ref{fig:fieldlines}). There
are some qualitative differences also between the homogeneous and
modulated wave field models (the top and middle panels). However, they
cannot be quantified on the basis of one realisation, and we will
study their difference by using the energetic particle simulations in
Section~\ref{sec:sep_propagation}.

 \begin{figure}
   \plotone{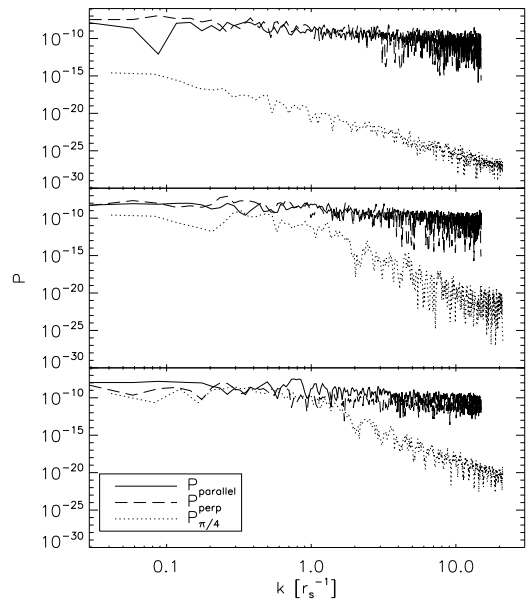}
   \caption{\label{fig:spectra}Power spectrum of the turbulence using
     the homogeneous (top panel), modulated wavefield (middle panel)
     and the random envelope turbulence model (bottom panel). In the
     latter two models, the envelope length is $L_p=4 L_c$ and density
     $\rho_p=4$.}
 \end{figure}

It is also important to determine how the turbulence spectrum is
affected by enveloping. This can be seen in Fig.~\ref{fig:spectra}
where we present a 2-dimensional Fourier transform of the turbulence,
for the homogeneous (top panel), the modulated wave field (middle
panel) and the random envelope turbulence model (bottom panel). We
show the spectrum for waves with the $k$ along the mean magnetic field
(solid line), along an axis perpendicular to it (dashed line), and the
along direction $45^\circ$ away from the mean magnetic field (dotted
line). It can be seen that for the homogeneous turbulence the wave
power resides in the parallel and perpendicular components only,
whereas the enveloping effectively isotropises the turbulence at
scales larger than the envelope length ($L_p=4 L_c$). In the random
envelope model, the perpendicular component of the spectrum (the
dashed line in bottom panel of Fig.~\ref{fig:spectra}) is reduced
compared to the two other models. This may have an influence on
energetic particle transport.

It should be noted that in this Section we have presented the
behaviour of the amplitudes, fieldlines and spectra for only one
realisation, and one region. Thus, the presented behaviour can be only
be considered as qualitative, as localised structures are apparent
even in homogeneous turbulence models \citep[e.g.][]{Chuychai2007}.

\subsection{Energetic Particle Propagation}\label{sec:sep_propagation}

\begin{figure}
   \plotone{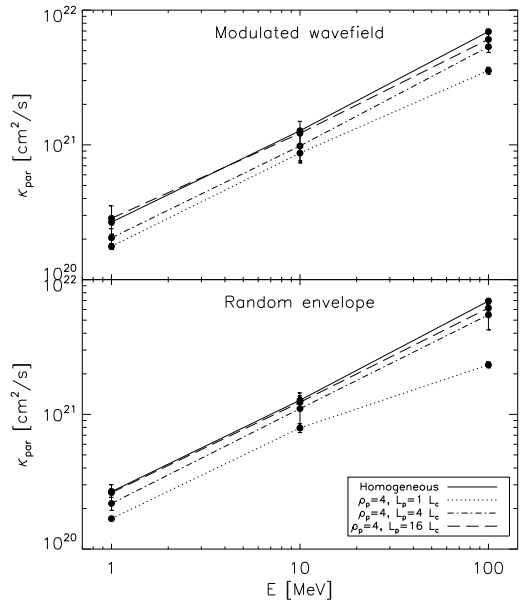}
   \caption{\label{fig:kappapar_ene}Parallel diffusion coefficient as a
     function of particle energy, for the modulated wavefield model
     (top panel) and for the random envelope model (bottom panel), for
     $\rho_p=4$ and four envelope lengths. The solid line represents
     the result for homogeneous turbulence. The turbulence amplitude is
     $B_1^2/B_0^2=1$.}
 \end{figure}

To study the influence of enveloping the turbulence on energetic
particle propagation, we have simulated several different scenarios,
varying the particle energy and the turbulence properties. In
Fig.~\ref{fig:kappapar_ene} we show the parallel diffusion
coefficients obtained from these runs, for protons of energies 1~MeV,
10~MeV and 100~MeV. The points represent the mean, and the
error-limits the standard deviation of the diffusion coefficients,
obtained from 10 turbulence realisations. The top panel shows the
modulated wave field results and the bottom one those of the random
envelope model.

In both panels of Fig.~\ref{fig:kappapar_ene}, the solid black line
depicts the diffusion coefficient obtained from runs with homogeneous
turbulence, given by Eq.~(\ref{eq:gjturb}), which are consistent with
the results of \citet{GiaJok1999}. The non-solid lines show the
parallel diffusion coefficient for enveloped turbulence, with
intermediate envelope density ($\rho_p=4$), and three envelope
lengths. The enveloping reduces the diffusion coefficients somewhat
for both the modulated wavefield model up to factor of 2 compared to
the homogeneous turbulence model (top panel) and the random envelope
model up to factor of 3 (bottom panel), and has a dependence on the
envelope length.

 \begin{figure}
   \plotone{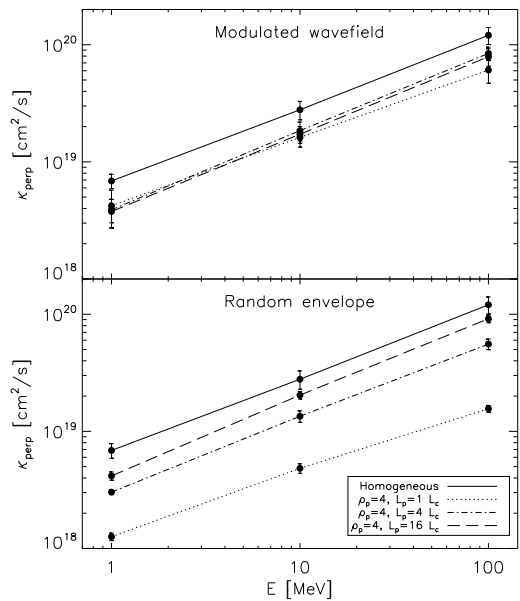}
   \caption{\label{fig:kappaperp_ene}Perpendicular diffusion
     coefficient as a function of energy, for the modulated wavefield
     model (top panel) and for the random envelope model (bottom
     panel). The turbulence amplitude is $ B_1^2/B_0^2=1$.}
 \end{figure}

For the perpendicular diffusion coefficient, the effect of the
enveloping is more varied. In Fig.~\ref{fig:kappaperp_ene}, we show
the variation of the perpendicular diffusion coefficient as a function
of energy for the same parameters as in Fig.~\ref{fig:kappaperp_ene}.
In the modulated wavefield model, the perpendicular diffusion
coefficient is reduced by factor of two compared to the homogeneous
turbulence result, and is independent of the envelope length. For the
random envelope model, the reduction is strongly dependent on the
envelope length, and the reduction in the diffusion coefficient
reaches factor of 8 for short envelopes.

 \begin{figure}
   \plotone{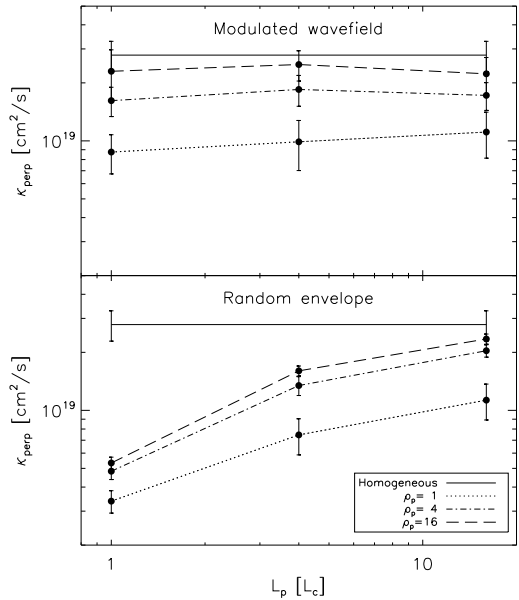}
   \caption{\label{fig:kappaperp_lencomp}Perpendicular diffusion
     coefficient as a function of the envelope length, for the
     modulated wavefield model (top panel) and for the random
     envelope model (bottom panel). The turbulence amplitude is $
     B_1^2/B_0^2=1$, and the particle energy 10~MeV.}
 \end{figure}

The effect of the enveloping parameters can be seen in more detail in
Fig.~\ref{fig:kappaperp_lencomp}, where we show the perpendicular
diffusion coefficient for three envelope densities, as a function of
the envelope length. For the modulated wavefield model (top panel),
there is no variation of the perpendicular diffusion with the envelope
length, within statistical errors of our study. There is a clear
dependence on the envelope density, however, with a reduction of the
diffusion coefficient up to factor 3 compared to the homogeneous
turbulence result, when the turbulence enveloping is tenuous
($\rho_p=1$ in top panel of Fig.~\ref{fig:kappaperp_lencomp}). Higher
envelope densities result in the diffusion coefficient approaching the
homogeneous turbulence values

When the phases, propagation directions and polarisations are
different from one envelope to another, the cross-field particle
propagation depends considerably on both enveloping parameters (bottom
panel of Fig.~\ref{fig:kappaperp_lencomp}). For a high-density,
long-envelope packeting, the limit of homogeneous turbulence is again
obtained. When the envelopes are shortened and the envelope density is
reduced, however, the reduction of the perpendicular diffusion
coefficient is nearly an order of magnitude.

\section{Discussion and Conclusions}

In this work, we have studied the effect of structured turbulence, as
opposed to homogeneous turbulence, on energetic particle
transport. We find that in the structured turbulence the particle
propagation is significantly inhibited, as compared to the propagation
in homogeneous turbulence of similar mean energy density of the
turbulent magnetic field.

The parallel diffusion can be seen to be affected strongly only when
the inhomogeneities are of small scales (the dotted curve in
Fig.~\ref{fig:kappapar_ene}). The reduction of the coefficient has
also a significant energy-dependence, with higher-energy particles
being affected more strongly. This is may be related to the
increasing resonant scales of the particles (with Larmor radius of
$0.4~r_\odot$ for 100 MeV protons in a 5~nT magnetic field). 

The effect of the structuring of the turbulence on the cross-field
propagation depends strongly on how the enveloping is performed. In
the modulated wave-field model, where only the amplitude of the
turbulence is modified, the cross-field propagation is sensitive only
to the envelope density. For short envelopes, it appears that the
effect of reduced field line wandering in the regions with
small-amplitude turbulence is not fully compensated by the increased
wandering in the higher-amplitude regions, but overall the cross-field
propagation is inhibited. With the increase of the envelope density,
the variation of the turbulence amplitudes approaches the homogeneous
description (see Fig.~\ref{fig:amplitude}), and the limit of
cross-field propagation in the homogeneous turbulence is approached
with increasing envelope density, as can be seen in the top panel of
Fig.~\ref{fig:kappaperp_lencomp}.

For the random envelope model, however, the effect of the enveloping
to the cross-field propagation is more varied, and stronger, and both
the length and the density affect the perpendicular diffusion
coefficient. Thus, the loss of phase coherence along the mean field
direction, which is the basic difference between the wavefield
modulation and random envelope models, affects significantly to the
cross-field propagation of energetic particles.

This effect can be understood qualitatively. Strong
  cross-field wandering of a field-line, as seen in the top panels of
  Fig.~\ref{fig:fieldlines}, can only take place due to persistent
  cross-field disturbance with little or no dependence on the
  coordinate along the background magnetic field. This is possible due
  to the 2D component of the turbulence: a 2D wave has a constant
  phase along the background magnetic field and the cross-field
  direction perpendicular to the $k$-vector. In the homogeneous and
  modulated wave-field model such waves are coherent throughout the
  simulation region (subject to disturbances from other waves),
  resulting in strong wandering of some field lines. In the random
  envelope model, however, this coherence is broken, and such
  wandering can remain coherent only when one envelope dominates the
  others.

Such breaking of the behaviour has natural justifications,
particularly in the heliospheric context. As discussed in the
Section~\ref{sec:introduction}, the heliospheric magnetic field, and
thus the turbulence, has its origins at the Sun, and the heliospheric
plasmas are known to be structured due to transient phenomena at the
Sun. Thus, it does not not seem feasible that the evolving corona
could produce structures with infinite phase coherence into the
heliospheric magnetic fields.

The non-linear evolution of turbulence also favours the evolution of
the turbulence in parallel direction. As waves propagate outward from
the Sun, they interact nonlinearly, resulting in cascade of energy
between different scales \citep[e.g.,][]{TuMarsch1995}. Although this
energy cascade is very anisotropic, towards small perpendicular
scales, the parallel scales are also affected. This can be understood
through the concept of critical balance, introduced by
\citet{GoSr1995}, where the turbulence separated by the critical scale
parallel to the mean field experience differing interactions due to
the evolution of the counterstreaming waves. This results in
scale-dependent anisotropy between the perpendicular and parallel
scales of the turbulence, which has been observed in the solar wind
turbulence \citep{Horbury2008}.

Overall, our results suggest that inhomogeneities within
heliospheric plasmas and the nonlinear behaviour of the turbulence may
significantly influence energetic particle transport, 
This has significant implications on the
studies of SEP origins and their propagation in the heliospheric
magnetic fields.

\acknowledgments

We acknowledge support from the UK Science and Technology Facilities
Council via standard grant ST/H002944/1 and a PhD studentship. 
Access to the University of Central Lancashire's High Performance
Computing Facility is gratefully acknowledged.

\appendix

\section{Calculation of the mean turbulence energy density} \label{sec:turbe}

For the random envelope model, the energy density given by
Eq.~(\ref{eqn:envEdens}) can be estimated analytically. As the magnetic
fields $\delta \mathbf{B}_i$ within different envelopes are not
correlated, we can consider the integral over terms with $i\neq j$
negligible. With this approximation, the integral can be written in
the form
$$
 U_{env} = \frac{1}{8\pi}\frac{1}{L_{max}S}\sum_{i=1}^{N_p} 
 \int_{0}^{L_{max}}\delta B^2_i\mathfrak{A}_i(z)^2\; dz\; dS.
$$
Using $\delta B^2(x,y,z)\approx\left<\delta b^2\right>$ on the scales
where $\mathfrak{A}_i$ changes appreciably, and ignoring the envelopes
that are only partly in the integration region, the integral results
in
$$U_{env} = \frac{3}{8} \frac{L_p}{L_{max}} U_{inf},$$
where $U_{inf}$ is the energy density of the turbulence given by
Eq.~(\ref{eq:gjturb}). Thus, the scaling factor for the enveloped
turbulence energy density is 
$$\frac{U_{inf}}{\sum_{N_p}U_{env}}=\frac{8}{3} \frac{L_{max}}{N_p
  L_p}
=\frac{8}{3 \rho_p}.$$
The magnetic field amplitude is scaled by the square root of this
scaling factor. We have verified the validity of this estimate through
numerical integration of $\delta B^2$. 

For the wavefield modulation model, the amplitudes can simply be added
together, as the wave phases and polarisations are the same in all
envelopes. At high envelope density, the sum approaches constant, thus
the scaling factor for the energy density is simply $4/\rho_p^2$. On
the limit of low envelope density, the envelopes do not overlap,
resulting in the same expression for the energy density as in the
random envelope model, and the scaling factor $8/(3\rho_p)$. However,
as the envelope densities in this study are in the intermediate range
between these asymptotes, we use numerical integration for determining
the scaling factor in the wavefield modulation model.

\bibliographystyle{apj}

\clearpage

\end{document}